\documentclass[aps,twocolumn,prl,groupedaddress,amsmath,amssymb]{revtex4}

\usepackage{graphicx}
\usepackage{dcolumn}
\usepackage{xcolor}%
\usepackage{booktabs}%
\usepackage{bm}
\usepackage{color, soul}

\begin{document}

\title{Cocktail effect on superconductivity in hexagonal high-entropy alloys}

\date{\today}

\author{Bin Liu$^{1}$\footnote{Email address: bliu0201@foxmail.com}}
\author{Wuzhang Yang$^{2,3,4}$}
\author{Guang-Han Cao$^{5}$}
\author{Zhi Ren$^{2,3}$\footnote{Email address: renzhi@westlake.edu.cn}}

\affiliation{$^{1}$Faculty of Materials Science and Engineering, Kunming University of Science and Technology, Kunming, Yunnan 650000, PR China}
\affiliation{$^{2}$School of Science, Westlake University, 18 Shilongshan Road, Hangzhou 310024, Zhejiang Province, PR China}
\affiliation{$^{3}$Institute of Natural Sciences, Westlake Institute for Advanced Study, 18 Shilongshan Road, Hangzhou 310024, Zhejiang Province, PR China}
\affiliation{$^{4}$Department of Physics, Fudan University, Shanghai 200433, PR China}
\affiliation{$^{5}$School of Physics, Zhejiang University, Hangzhou 310058, PR China}

\begin{abstract}
We report the study of the cocktail effect on superconductivity in high-entropy alloys (HEAs), using hexagonal close-packed HEAs as a prototype system.
Compared with the compositional averages of the constituent elements, the superconducting transition temperature $T_{\rm c}$ is enhanced by from a factor of $\sim$2 to over one order of magnitude.
This $T_{\rm c}$ enhancement correlates with the reduction in the Debye temperature, underlining the importance of phonon softening in triggering the cocktail effect.
Furthermore, we show that the $T_{\rm c}$ in these HEAs is governed by the average phonon frequency and electron-phonon coupling strength, the latter of which scales linearly with the inverse HEA molecular weight and is progressively weakened with increasing mixing entropy.
Our study paves the way toward quantitative understanding of the superconductivity in HEAs.
\end{abstract}

\maketitle
\maketitle

\section{1. Introduction}
Over the past decade, a new class of compositionally complex alloys developed based on the entropy concept, the high-entropy alloys (HEAs), have attracted much attention \cite{ref1,ref2,ref3,ref4}.
These emerging alloys are made up of four or more principle elements with equimolar or nearly equimolar ratios.
This chemical complexity leads to enhanced mixing entropy, which overcomes the enthalpies of compound formation.
As a consequence, HEAs exist in a single solid-solution phase and are often regarded as metallic glasses on an ordered lattice.
Compared with traditional alloys based on one or two principle elements, HEAs often exhibit superior mechanical and physical properties,
such as high hardness \cite{ref5}, high thermal stability \cite{ref6}, excellent corrosion resistance \cite{ref7}, complex magnetism \cite{ref8}, and robust superconductivity against disorder \cite{ref9}.
\begin{table*}
	\caption{Physicochemical and physical parameters of the hexagonal HEA superconductors. The data for Mo$_{0.12}$Re$_{0.88}$ and MoReRu alloys are also included for comparison.}
	\renewcommand\arraystretch{1.4}
	\begin{tabular}{p{4.5cm}<{\centering}p{0.8cm}<{\centering}p{0.8cm}<{\centering}p{0.8cm}<{\centering}p{0.8cm}<{\centering}p{0.8cm}<{\centering}p{1.7cm}<{\centering}p{1.7cm}<{\centering}p{0.8cm}<{\centering}p{0.8cm}<{\centering}p{0.8cm}<{\centering}p{1.2cm}<{\centering}}\\
		\hline
		Composition & $\delta$(\%) & $\Delta$$\chi$ & VEC &  $T_{\rm c}$ (K) &  $\overline{T}_{\rm c}$ (K)  & $\gamma$ (mJ/molK$^{2})$ & $\overline{\gamma}$ (mJ/molK$^{2})$ & $\Theta_{\rm D}$ (K) &$\overline{\Theta}$$_{\rm D}$ (K) &$\lambda_{\rm ep}$  & $\Delta$$S_{\rm mix}$ \\
		\hline
        Mo$_{50}$Ru$_{20}$Rh$_{15}$Pd$_{15}$$^{[13]}$& 1.763 &0.052 & 7.45&5.0&0.15&$-$ & $-$& $-$& $-$& $-$ & 1.237$R$  \\
        Mo$_{40}$Ru$_{40}$Rh$_{10}$Pd$_{10}$$^{[13]}$& 1.997 &0.0354& 7.5 &3.0&0.32& $-$&$-$& $-$& $-$ & $-$ & 1.193$R$  \\
        MoReRuRh$^{[14]}$& 1.741 &0.142& 7.5 &2.5&0.58&$-$&$-$ & $-$& $-$& $-$ & 1.386$R$   \\
        (MoReRuRh)$_{0.95}$Ti$_{0.05}$$^{[14]}$& 2.104 &0.187& 7.33&3.6&0.57& $-$&$-$& $-$& $-$& $-$  & 1.515$R$   \\
        (MoReRuRh)$_{0.9}$Ti$_{0.1}$$^{[14]}$& 2.076 &0.224& 7.15&4.7&0.69& $-$&$-$& $-$& $-$& $-$  & 1.573$R$   \\
        Nb$_{10}$Mo$_{35}$Ru$_{35}$Rh$_{10}$Pd$_{10}$$^{[15]}$& 2.134 &0.181 & 7.3&5.58&1.09& 3.35& 4.27& 348& 424&0.63 & 1.426$R$ \\
        Nb$_{15}$Mo$_{32.5}$Ru$_{32.5}$Rh$_{10}$Pd$_{10}$$^{[15]}$& 2.105 &0.184& 7.2&6.19&1.52& 3.27& 4.58& 328& 463&0.66  & 1.476$R$   \\
        Nb$_{20}$Mo$_{30}$Ru$_{30}$Rh$_{10}$Pd$_{10}$$^{[15]}$& 2.076 &0.187& 7.1&6.10&1.95& 2.68& 3.88& 324&449&0.66 & 1.505$R$   \\
        Nb$_{5}$Mo$_{35}$Re$_{15}$Ru$_{35}$Rh$_{10}$$^{[16]}$& 2.074 &0.157& 7.1 &7.54&0.89& 3.70&3.21& 338&502 &0.70 & 1.40$R$  \\
        Nb$_{5}$Mo$_{30}$Re$_{20}$Ru$_{35}$Rh$_{10}$$^{[16]}$& 2.048 &0.160& 7.15&6.69&0.97& 3.93& 3.22& 467& 501&0.62  & 1.431$R$   \\
        Nb$_{5}$Mo$_{25}$Re$_{25}$Ru$_{35}$Rh$_{10}$$^{[16]}$& 1.997 &0.163& 7.2&6.51&1.06& 4.15& 3.24& 393& 500&0.64  & 1.441$R$   \\
        Nb$_{5}$Mo$_{20}$Re$_{30}$Ru$_{35}$Rh$_{10}$$^{[16]}$& 1.985 &0.165& 7.25&5.46&1.14& 3.32& 3.76& 378& 499&0.61  & 1.431$R$   \\
        (MoReRu)$_{0.916}$(PdPt)$_{0.084}$$^{[17]}$& 1.698 &0.135& 7.25&8.17&0.69& 3.55& 3.11& 364& 486&0.70  & 1.353$R$   \\
        (MoReRu)$_{0.834}$(PdPt)$_{0.166}$$^{[17]}$& 1.636 &0.135& 7.5&4.91&0.63& 3.01& 3.57& 393& 466&0.59  & 1.481$R$   \\
        (MoReRu)$_{0.75}$(PdPt)$_{0.25}$$^{[17]}$& 1.570 &0.134& 7.75&2.22&0.58& 2.77& 4.04& 411& 445&0.49  & 1.56$R$   \\
        (MoReRu)$_{0.66}$(PdPt)$_{0.34}$$^{[17]}$& 1.498 &0.133& 8.0&1.64&0.52& 2.79& 4.55& 398& 421&0.47  & 1.60$R$   \\
        MoReRuRhPt$^{[18]}$ & 1.57 &0.140& 8.0&0.82&0.48& $-$&$-$&$-$&$-$&$-$  & 1.61$R$   \\
        MoReRuIrPt$^{[18]}$& 1.47 &0.130& 8.0&1.06&0.48& $-$&$-$&$-$&$-$&$-$  & 1.61$R$   \\
        MoReRuRhPd$^{[18]}$& 1.20 &0.130& 8.0&1.38&0.48& $-$&$-$&$-$&$-$&$-$  & 1.61$R$  \\
        MoReRuIrPd$^{[18]}$& 1.10 &0.117& 8.0&1.61&0.48&$-$&$-$&$-$&$-$&$-$  & 1.61$R$   \\
        NbMoReRuRhPt$^{[18]}$\footnote{Sample contains a secondary hexagonal phase.} & 2.42 &0.246& 7.5 &2.95&1.88& $-$&$-$&$-$&$-$&$-$  & 1.79$R$   \\
        NbMoReRuIrPt$^{[18]}$$^{a}$ & 2.32 &0.236& 7.5 &2.84&1.88& $-$&$-$&$-$&$-$&$-$  & 1.79$R$   \\
        Ru$_{0.35}$Os$_{0.35}$Mo$_{0.1}$W$_{0.1}$Zr$_{0.1}$$^{[19]}$ & 4.26 &0.278& 7.2&2.90&0.50& 2.42& 2.63& 357& 504&0.52  & 1.43$R$   \\
        Mo$_{35}$W$_{10}$Re$_{20}$Ru$_{30}$Pd$_{5}$$^{[20]}$ & 1.84 &0.133& 7.0&8.32&0.51& 3.29& 2.85& 277&490&0.79  & 1.43$R$   \\
        MoReRu$^{[14]}$ & 1.84 &0.133& 7.0&9.25&0.74& 4.23& 2.64& 339&507&0.76  & 1.10$R$  \\
        Mo$_{0.12}$Re$_{0.88}$$^{[31]}$ & $-$ &$-$& 6.88&7.45&1.50& 3.80& 2.41& 403&452&0.66  & $-$   \\
		\hline
	\end{tabular}
	\label{Table1}
\end{table*}

In principle, HEAs are anticipated to possess four core effects, including a high-entropy effect for thermodynamics, a lattice distortion effect for structure, a sluggish diffusion effect for kinetics, and a cocktail effect for properties \cite{ref2,ref3,ref4}.
Especially, the cocktail effect means that the elements that make up the HEAs work in synergy with each other to enable better performance than the compositional average.
This has indeed been demonstrated for the mechanical, catalytic, electrochemical and hydrogen storage properties \cite{ref10,ref11}. In contrast, the cocktail effect on the physical properties has been poorly explored.

Hexagonal close-packed structure is one of the prototype structures for HEAs \cite{ref12}. Although the hexagonal HEAs accounts for only $\sim$1\% of the total HEAs, those based on 4$d$/5$d$ elements are known for their proneness to become a superconductor.
So far, nearly thirty different superconducting compositions have been found in this family \cite{ref13,ref14,ref15,ref16,ref17,ref18,ref19,ref20}, and
their $T_{\rm c}$ varies by one order of magnitude up to $\sim$8.3 K, as listed in Table 1.
Notably, quite a few of these HEAs, such as Mo-Ru-Rh-Pd \cite{ref13}, Mo-Re-Ru-Rh \cite{ref14}, Mo-Re-Ru-Pd-Pt \cite{ref17}, and Mo-W-Re-Ru-Pd \cite{ref20}, have $T_{\rm c}$ values much higher than those of the corresponding constituent elements.
This clearly violates the rule of mixtures and signifies enhanced superconducting pairing induced by the elemental makeup. In this regard, the elucidation of its mechanism not only helps to better understand the cocktail effect and but also may provide clues to achieve a higher $T_{\rm c}$. However, to our knowledge, no such study has been reported to date.
\begin{figure*}
	\includegraphics*[width=16.5cm]{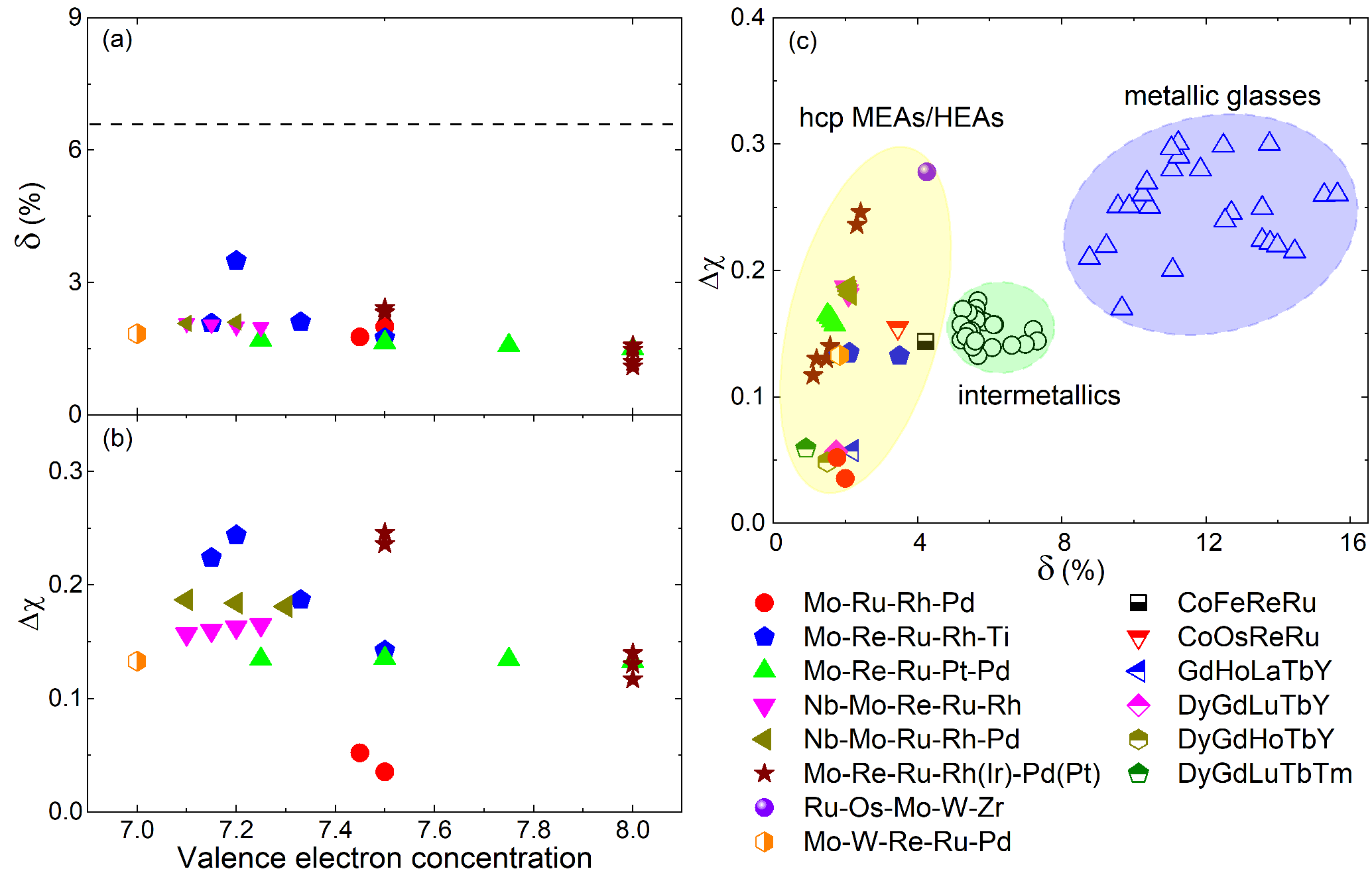}
	\caption{
		(a, b) VEC dependencies of $\delta$ and $\Delta$$\chi$, respectively, for the hexagonal HEA superconductors.
        In panel (a), the horizontal dashed line denotes the threshold for the formation of single-phase HEAs.
        (c) $\delta$ versus $\Delta$$\chi$ plot of the hexagonal compositionally complex alloy superconductors. The data for rare-earth hexagonal HEAs, typical intermetallics, and metallic glasses are also included for comparison.
        The regions corresponding to different phases are highlighted in different colors.
	}
	\label{fig1}
\end{figure*}

Here we address this issue by comparing the experimental $T_{\rm c}$, Sommerfeld coefficient $\gamma$, and Debye temperature $\Theta_{\rm D}$ values with the compositionally averaged counterparts $\overline{T}_{\rm c}$, $\overline{\gamma}$ and $\overline{\Theta}_{\rm D}$.
It is found that, with varying valence electron concentration (VEC), the increase in $T_{\rm c}$/$\overline{T}_{\rm c}$ is concomitant with the decrease in $\Theta_{\rm D}$/$\overline{\Theta}_{\rm D}$.
The $\lambda_{\rm ep}$ dependence of $T_{\rm c}$ and its affecting factors are examined, and the implications of these results are discussed.

\section{2. Phase stability}
In the hexagonal close-packed lattice, there is only one crystallographic site and hence a solid-solution phase is naturally expected.
To confirm this, we calculate the atomic size difference $\delta$ and electronegativity difference $\Delta$$\chi$ according to the following equations \cite{ref21,ref22},
\begin{equation}
\delta = \sqrt{\sum^{n}_{i=1}c_{i}(1-r_{i}/\sum^{n}_{i=1}c_{i}r_{i})^{2}},
\end{equation}
and
\begin{equation}
\Delta\chi = \sqrt{\sum^{n}_{i=1}c_{i}(\chi_{i}-\sum^{n}_{i=1}c_{i}\chi_{i})^{2}},
\end{equation}
where $n$ is the number of the constituent elements, and $c_{i}$, $r_{i}$ and $\chi_{i}$ are the molar fraction, atomic radius and electronegativity for the $i$th constituent element, respectively (variables are the same hereafter).
The calculated results are listed in Table I, and $\delta$ as well as $\Delta$$\chi$ are plotted as a function of VEC in Figs. 1(a) and (b), respectively. Here VEC is given as
\begin{equation}
\rm VEC = \sum^{n}_{i=1}\it c_{\it i} (\rm VEC)_{\it i},
\end{equation}
where (VEC)$_{i}$ is the valence electron concentration for the $i$th constituent element.
The superconducting hexagonal HEAs exist over a wide range of 7.0 $\leq$ VEC $\leq$ 8.0, yet their $\delta$ and $\Delta$$\chi$ fall in the narrow windows of 1.1$-$3.5\% and 0.04$-$0.25, respectively.
Note that the $\delta$ values are well below the threshold of 6.6\% for the formation of single-phase HEAs.
Fig. 1(c) shows the $\delta$ plotted as a function of $\Delta$$\chi$, together with data of hexagonal rare earth HEAs, intermetallic compounds and metallic glasses \cite{ref23,ref24,ref25} for comparison.
As expected, the data of all the hexagonal HEAs overlap and are located within the region of 0.9\% $\leq$ $\delta$ $\leq$ 4.2\% and 0.035 $\leq$ $\Delta\chi$ $\leq$ 0.278,
which are separated from those of the intermetallic compounds and metallic glasses with larger $\delta$ values.
\begin{figure}
	\includegraphics*[width=7.5cm]{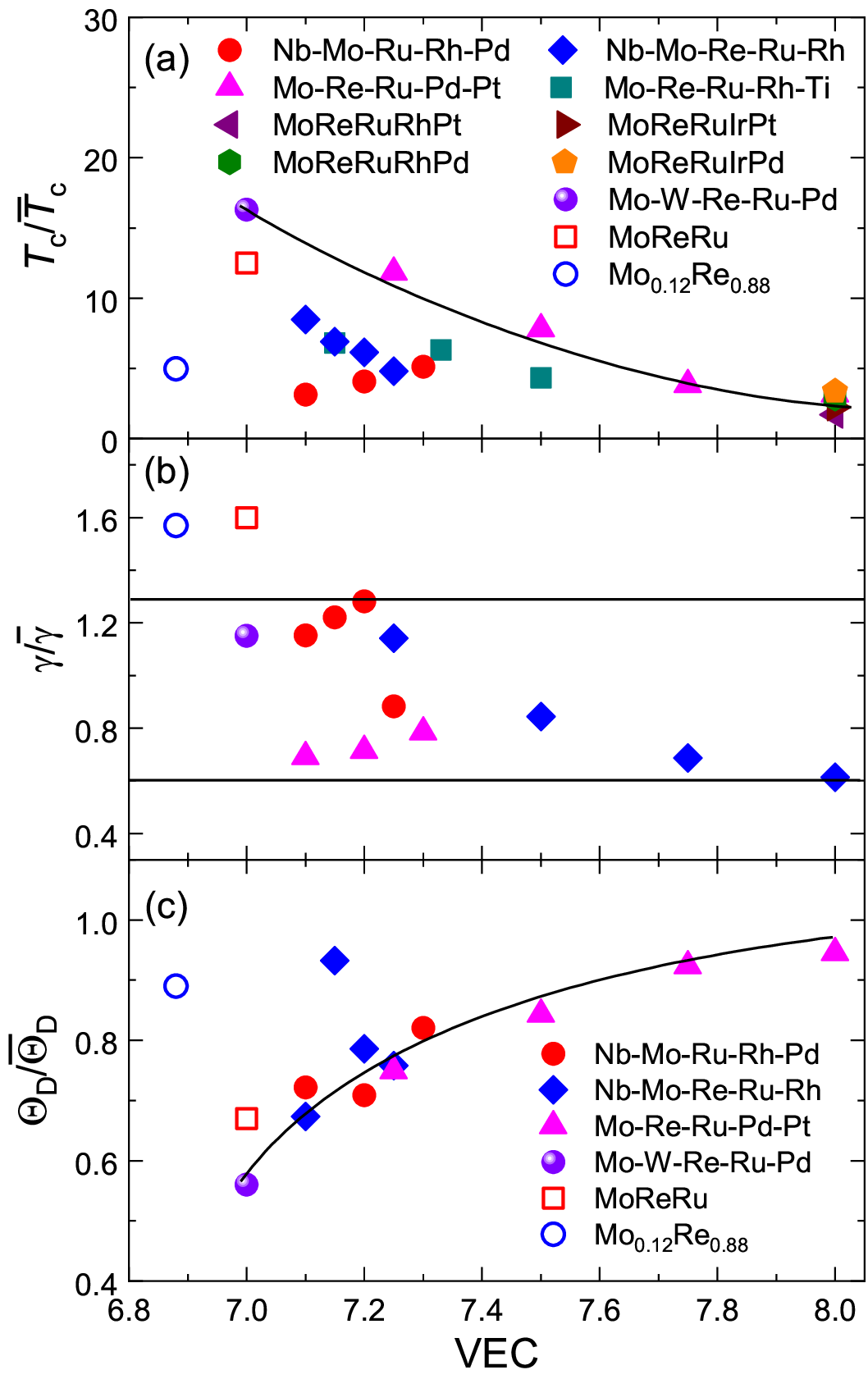}
	\caption{
		(a-c) VEC dependencies of $T_{\rm c}$/$\overline{T}_{\rm c}$, $\gamma$/$\overline{\gamma}$, and $\Theta_{\rm D}$/$\overline{\Theta}_{\rm D}$, respectively, for the hexagonal HEA superconductors.
        The data for binary and ternary alloys are also included for comparison and the solid lines are a guide to the eyes.
	}
	\label{fig2}
\end{figure}

\section{3. Cocktail effect on $T_{\rm c}$, $\gamma$ and $\Theta_{\rm D}$}
The magnitude of the cocktail effect is elucidated by examining the ratios of $T_{\rm c}$/$\overline{T}_{\rm c}$, $\gamma$/$\overline{\gamma}$, and $\Theta_{\rm D}$/$\overline{\Theta}_{\rm D}$, which are plotted as a function of VEC in Figs. 2(a-c), respectively.
Following Ref. [\citenum{ref26}], the compositionally-averaged $\overline{T}_{\rm c}$, $\overline{\gamma}$ and $\overline{\Theta}_{\rm D}$ are calculated using the following equation
\begin{equation}
\overline{X} = \sum^{n}_{i=1}c_{i}X_{i},
\end{equation}
where $X_{i}$ is the parameter value of the $i$th element (see Table II).
The experimental $T_{\rm c}$, $\gamma$ and $\Theta_{\rm D}$ values for the HEAs and individual elements are taken from Refs. [13-20] and Ref. [\citenum{ref38,ref40}], respectively.
At VEC = 7.0, the $T_{\rm c}$/$\overline{T}_{\rm c}$ ratio reaches $\sim$15; this values is much larger than that found in body-centered cubic (bcc) HEA superconductors \cite{ref27} and indicates a strong cocktail effect.
With increasing VEC, the $T_{\rm c}$/$\overline{T}_{\rm c}$ ratio decreases but remains above 2 up to VEC = 8.
As for $\gamma$/$\overline{\gamma}$ and $\Theta_{\rm D}$/$\overline{\Theta}_{\rm D}$, only the data for the series of Nb-Mo-Ru-Rh-Pd \cite{ref15}, Nb-Mo-Re-Ru-Rh \cite{ref16}, Mo-Re-Ru-Pd-Pt \cite{ref17} and Mo-W-Re-Ru-Pd \cite{ref20} HEAs are available.
However, since their VECs cover the range of 7.0 to 8.0, the results should be considered as representative.
Contrary to those of $T_{\rm c}$/$\overline{T}_{\rm c}$, the $\gamma$/$\overline{\gamma}$ ratios are all located near one with the deviation less than 30\%,
which are smaller than that observed in the bcc HEA superconductor \cite{ref27}.
Instead, the $\Theta_{\rm D}$/$\overline{\Theta}_{\rm D}$ increases with increasing VEC; this is in the opposite trend to that of $T_{\rm c}$/$\overline{T}_{\rm c}$.
At the highest VEC of 8.0, the $\Theta_{\rm D}$/$\overline{\Theta}_{\rm D}$ ratio ($\sim$0.95) is very close to one, indicating that the rule of mixtures applies well in the case.
As the decrease of VEC, however, the $\Theta_{\rm D}$/$\overline{\Theta}_{\rm D}$ ratio decreases by $\sim$40\% to 0.57.
The concomitant reduction in $\Theta_{\rm D}$/$\overline{\Theta}_{\rm D}$ and enhancement in $T_{\rm c}$/$\overline{T}_{\rm c}$ is reminiscent of the irradiation study on Mo$_{5}$Ge$_{3}$ \cite{ref28}.
While non-irradiated Mo$_{5}$Ge$_{3}$ is normal down to 1.6 K, superconductivity below 3.3 K is observed after irradiation and attributed to disorder-induced phonon softening.
According to McMillan \cite{ref29}, the $\lambda_{\rm ep}$ can be expressed as
\begin{equation}
\lambda_{\rm ep} \equiv \frac{N(0)\langle I^{2}\rangle}{M\langle\omega^{2}\rangle},
\end{equation}
where $N$(0) is the bare density of states at the Fermi level, $M$ is the molecular mass, and $\langle I^{2}\rangle$ and $\langle\omega^{2}\rangle$ are the averaged electron-phonon matrix elements and phonon frequencies, respectively.
After irradiation, the $N$(0) of Mo$_{5}$Ge$_{3}$ remains almost unaffected while its $\Theta_{\rm D}$ decreases significantly from 377 to 320 K \cite{ref28}.
Since $\Theta_{\rm D}$ is positively correlated with $\langle\omega^{2}\rangle$, this leads to a significant increase in $\lambda_{\rm ep}$.
The high-entropy alloying is similar to radiation in inducing strong atomic disorder and hence phonon softening, which leads to the cocktail of superconductivity properties in the hexagonal HEAs.
\begin{table}
	\caption{Parameters of the constituent elements for the hexagonal HEA superconductors. The data are taken from Ref. [\citenum{ref38,ref40}].}
	\renewcommand\arraystretch{1.4}
	\begin{tabular}{p{1.8cm}<{\centering}p{1.2cm}<{\centering}p{1.2cm}<{\centering}p{2.3cm}<{\centering}p{1.2cm}<{\centering}}\\
		\hline
		Element &  VEC &  $T_{\rm c}$ (K) &  $\gamma$ (mJ/molK$^{2})$ & $\Theta_{\rm D}$ (K) \\
		\hline
        Zr& 4.0 &0.7 & 3.03&310  \\
        Nb& 5.0 &8.8 & 8.82&250  \\
        Mo& 6.0 &0.05& 2.11 &470\\
        W& 6.0 &0.06& 1.21 &405  \\
        Re& 7.0 &1.7& 2.45&450  \\
        Ru& 8.0 &0.47& 3.35&600   \\
        Os& 8.0 &0.71& 2.35&500   \\
        Rh& 9.0 &0.09 & 4.89&478\\
        Ir& 9.0 &0.1& 3.14&420  \\
        Pd& 10.0 &0.1& 9.9&299   \\
        Pt& 10.0 &0.1& 6.63 &221 \\

		\hline
	\end{tabular}
	\label{Table1}
\end{table}

To corroborate the presence of cocktail effect, we have also calculated the $T_{\rm c}$/$\overline{T}_{\rm c}$, $\gamma$/$\overline{\gamma}$, and $\Theta_{\rm D}$/$\overline{\Theta}_{\rm D}$ of hexagonal Mo$_{0.12}$Re$_{88}$ \cite{ref41} and MoReRu \cite{ref14} alloys.
Compared with the HEAs with (nearly) the same VEC, the $T_{\rm c}$/$\overline{T}_{\rm c}$ ratios are indeed smaller while the $\Theta_{\rm D}$/$\overline{\Theta}_{\rm D}$ ratios are larger for the binary and ternary alloys, indicating a weakened cocktail effect.
For dilute alloys, the $T_{\rm c}$ is typical determined by the component with the highest $T_{\rm c}$.
However, as all the components approach equimolar ratios, this is no longer case, even for binary alloys.
For example, while Nb has the highest $T_{\rm c}$ among elements, the $T_{\rm c}$ of Nb$_{1-x}$V$_{x}$ and Ta$_{1-x}$V$_{x}$ alloys with $x$ near 0.5 are lower than those of both the end members \cite{ref39}.
Also in the case of the Nb-Mo-Ru-Rh-Pd HEAs, increasing the Nb content above 15\% results in the suppression of $T_{\rm c}$ \cite{ref15}.
Hence, it is the synergistic effect of all constituent elements rather than individual one that is responsible for the enhancement of $T_{\rm c}$ compared with the compositional average in hexagonal HEAs.

\section{4. Dependence of $T_{\rm c}$ on $\lambda_{\rm ep}$}
In Fig. 3, we plot the $T_{\rm c}$ data against $\lambda_{\rm ep}$ and $\gamma$ for all available cases, in which the $\lambda_{\rm ep}$ data are taken from Refs. [15-17,19,20].
In contrast with the scattered $\gamma$ dependence of $T_{\rm c}$ (see the inset of Fig. 3), the $T_{\rm c}$ data almost collapse on a single curve as a function of $\lambda_{\rm ep}$.
At first glance, this "universal" behavior may not be surprising since the $\lambda_{\rm ep}$ is calculated from $T_{\rm c}$ and $\Theta_{\rm D}$ using the inverted McMillan formula.
Nonetheless, this allows us to estimate the average logarithmic phonon frequency $\langle\omega_{\rm log}\rangle$ using the Allen-Dynes formula \cite{ref31}
\begin{equation}
T_{\rm c} = \frac{\langle\omega_{\rm log}\rangle}{1.20}\rm exp[\frac{-1.04(1+\lambda_{\rm ep})}{\lambda_{\rm ep}-\mu^{\ast}(1+0.62\lambda_{\rm ep})}],
\end{equation}
where $\mu^{\ast}$ is the Coulomb repulsion pseudopotential.
As can be seen, the $\lambda_{\rm ep}$ dependence of $T_{\rm c}$ can indeed be well reproduced with $\langle\omega_{\rm log}\rangle$ = 235 K and $\mu^{\ast}$ = 0.11.
Note that the assumption of this constant $\mu^{\ast}$ is reasonable since it is less important relative to $\lambda_{\rm ep}$ and lies within the range of 0.1 to 0.13 for transition metals \cite{ref29}.
Meanwhile, the $\langle\omega_{\rm log}\rangle$ value is well comparable with hexagonal Re metal (227 K) \cite{ref32} and its binary alloys Nb$_{0.18}$Re$_{0.82}$ (264 K) \cite{ref33}, Mo$_{1-x}$Re$_{x}$ (175-250 K)\cite{ref34}, suggesting that the $\langle\omega_{\rm log}\rangle$ is not significantly affected by the mixing entropy.
The overall results are reminiscent of those found in ternary Tl-Pb-Bi alloys \cite{ref35}, where the $\langle\omega_{\rm log}\rangle$ is found not to change markedly while the $T_{\rm c}$ varies appreciably.
However, the $\lambda_{\rm ep}$ of the Tl-Pb-Bi alloys were determined from tunneling experiments \cite{ref35}, which calls for similar studies on the hexagonal HEA superconductors to substantiate the $\lambda_{\rm ep}$-dependent $T_{\rm c}$ behavior.
\begin{figure}
	\includegraphics*[width=8.5cm]{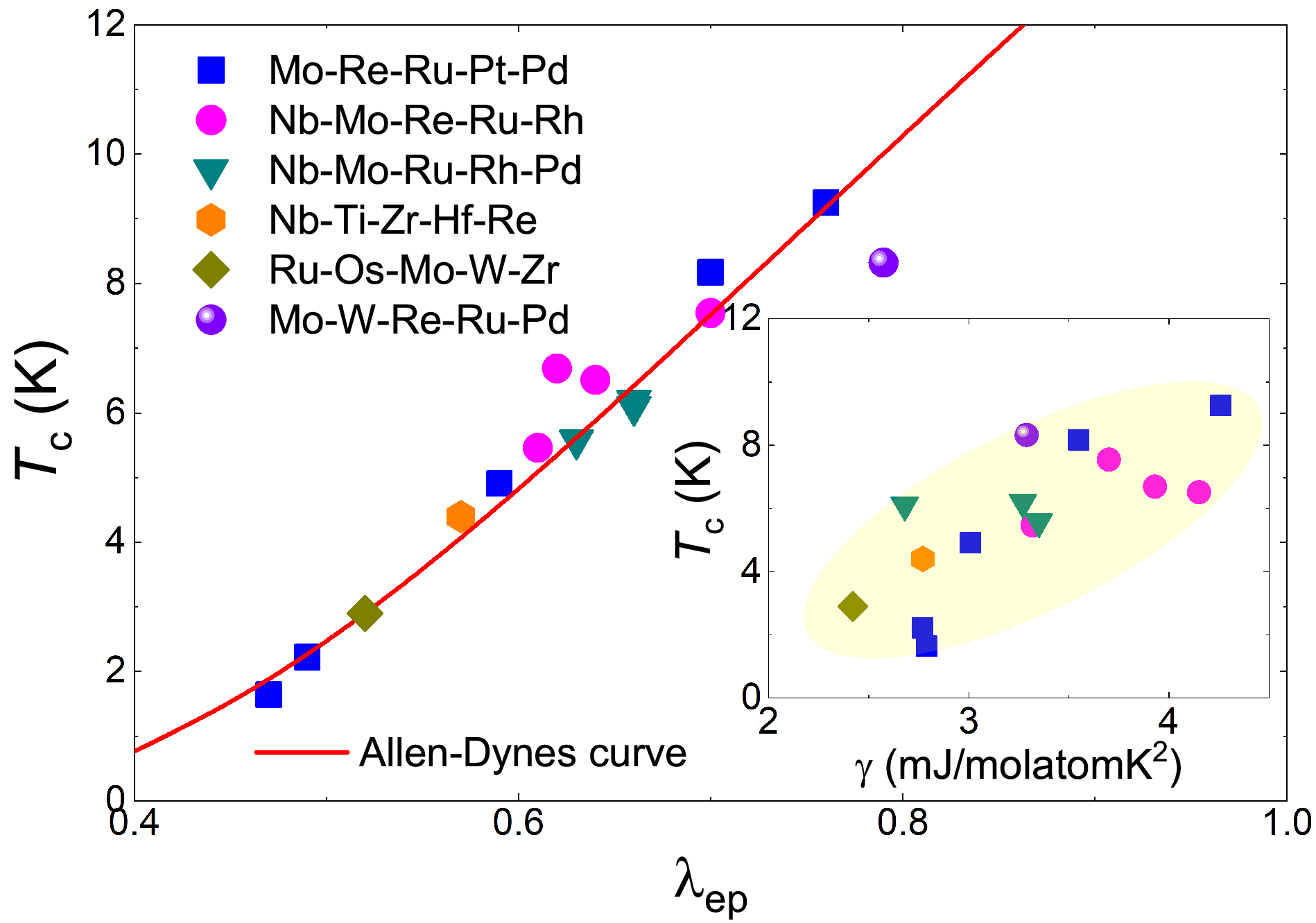}
	\caption{
	Available $\lambda_{\rm ep}$ dependence of $T_{\rm c}$ for the hexagonal HEA superconductors.
    The solid line denotes a fit to the data by the Allen-Dynes formula (see text).
    The inset shows the $T_{\rm c}$ data plotted as a function of $\gamma$.
    The shaded area is a guide to the eyes.
	}
	\label{fig2}
\end{figure}

\section{5. Affecting factors on $\lambda_{\rm ep}$}
We now discuss the factors that affect $\lambda_{\rm ep}$ based on equation (5).
For the Tl-Pb-Bi alloys \cite{ref35}, it has been shown that the product $N(0)$$\langle I^{2}\rangle$ is not constant throughout the series and the change in $\lambda_{\rm ep}$ is determined primarily by the variation of $N(0)$.
In Fig. 4(a), we plot the $\lambda_{\rm ep}$ against $\gamma$/(1+$\lambda_{\rm ep}$)$\propto$ $N(0)$ for the hexagonal HEA superconductors.
One can see that $\lambda_{\rm ep}$ is at most weakly dependent on $N(0)$, even for the same elemental combination.
As noticed by McMillan \cite{ref29}, the $N(0)$$\langle I^{2}\rangle$ remains constant to within 50\% for the transition metals V, Nb, Ta, Mo, W although their $N(0)$ and $\langle I^{2}\rangle$ vary by nearly one order of magnitude.
Such a scenario is more plausible for the hexagonal HEA superconductors considering their constituent elements.
Besides, $\langle\omega^{2}\rangle$ should not change much given the universal $\langle\omega_{\rm log}\rangle$.
Taken together, the $\lambda_{\rm ep}$ is then expected to scale with the inverse molecular weight of the alloys, $M_{\rm alloy}$.
As illustrated in Fig. 4(b), a linear dependence of $\lambda_{\rm ep}$ on 1/$M_{\rm alloy}$ is observed for all the series of Mo-Re-Ru-Rh-Pd, Nb-Mo-Re-Ru-Rh, and Nb-Mo-Ru-Rh-Pd HEAs.
The slopes for the former two series are comparable but considerably steeper than that of the latter one.
In passing, the data points of the Ru-Os-Mo-W-Zr and Mo-W-Re-Ru-Pd HEAs fall on the gentle and steep slopes, respectively.

\begin{figure}
	\includegraphics*[width=7.5cm]{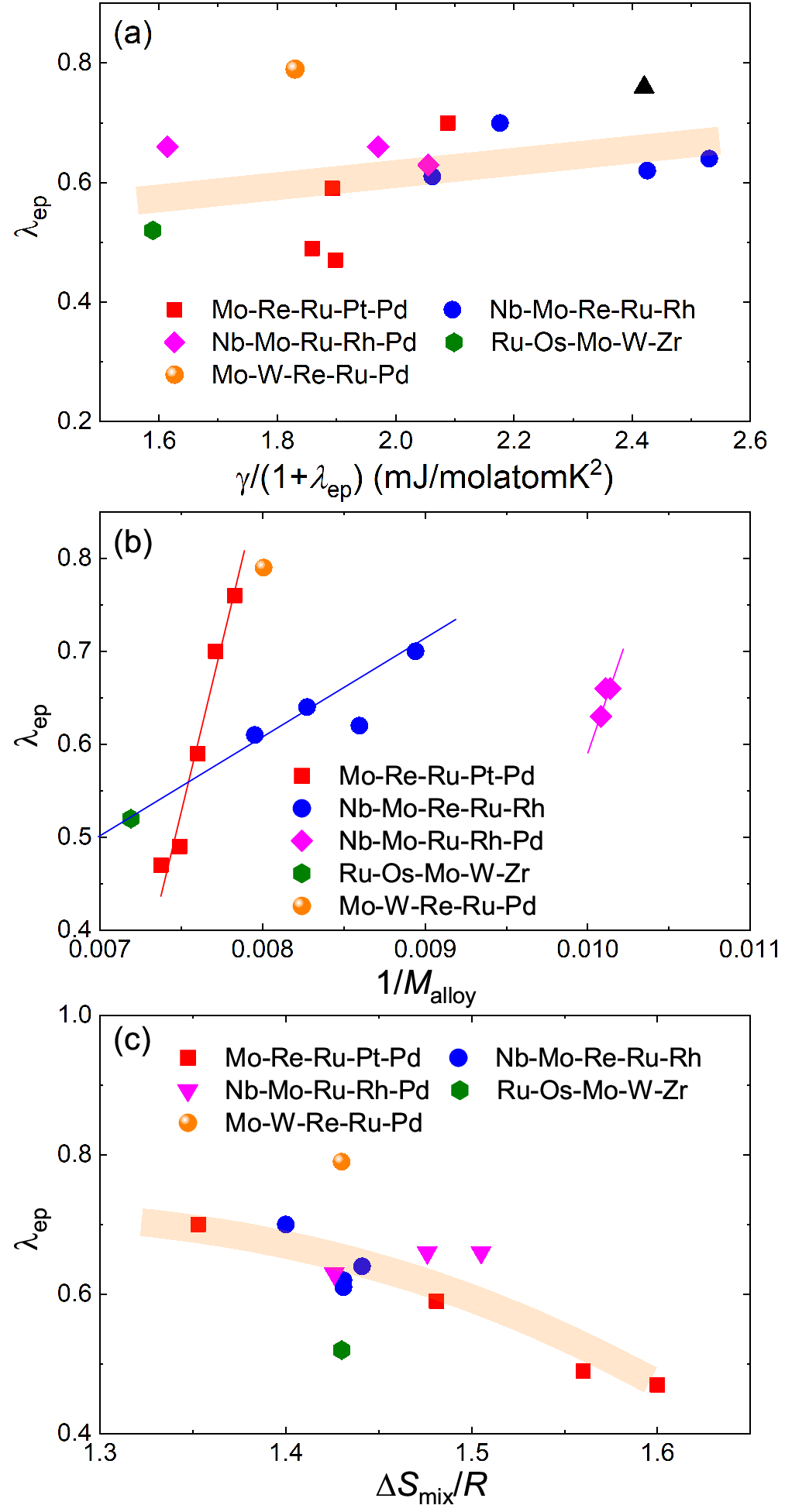}
	\caption{
	(a-c) Available $\lambda_{\rm ep}$ data plotted as functions of $\gamma$/(1+$\lambda_{\rm ep}$), 1/$M_{\rm alloy}$ and $\Delta$$S_{\rm mix}$, respectively, for the hexagonal HEA superconductors.
     The thin and thick lines are guides to the eyes.
	}
	\label{fig2}
\end{figure}

A salient feature for the hexagonal HEA superconductors is their large mixing entropy $\Delta$$S_{\rm mix}$ given by the equation
\begin{equation}
\Delta S_{\rm mix} = -R\sum^{n}_{i=1}\it c_{\it i}\rm ln \it c_{\it i},
\end{equation}
where $R$ is the gas constant.
In this regard, the $\Delta$$S_{\rm mix}$ dependence of $\lambda_{\rm ep}$ deserves scrutiny and is displayed in Fig. 4(c).
With increasing $\Delta$$S_{\rm mix}$ from $\sim$1.43$R$ to $\sim$1.60$R$, $\lambda_{\rm ep}$ decreases by more than 30\% and the slope becomes steeper, exhibiting a downward curvature.
Hence the electron-phonon interaction is reduced progressively as the atomic disorder gets stronger.
In HEA systems, both the electronic and phonon bands are expected to be broadened \cite{ref36,ref37}, which smears our the electronic and phononic density states at the Fermi level.
Since the $T_{\rm c}$ of hexagonal HEA superconductors is mainly governed by the $\lambda_{\rm ep}$ rather than the $\gamma$, the broadening of the phonon band should play a more important role than that of the electronic one, consistent with the above result.

We now propose several routes to increase the $\lambda_{\rm ep}$ in hexagonal HEA superconductors.
According to Fig. 4(b), one straightforward way will be to decrease the $M$ by incorporating elements with a lighter atomic mass.
Potential candidates include group IIIB elements Sc and Y, both of which can exist in the hexagonal structure.
In addition, group IIIA elements Al and Ga, which are known to be useful additives, are also worth trying.
On the other hand, increasing the the product $N(0)$$\langle I^{2}\rangle$ provides an alternative approach.
This implies an enhancement of the electron-ion interaction and can be achieved by carefully choosing the constituent elements for the HEAs.
Especially, the group VIIB element Re appear to be an indispensable component to achieve a large $N(0)$$\langle I^{2}\rangle$.
In either case, the $\Delta$$S_{\rm mix}$ should be kept not too large so that its influence on the $\lambda_{\rm ep}$ is minimized.

\section{6. Conclusion}
In summary, we have studied the cocktail effect in hexagonal HEA superconductors.
It is found that the $T_{\rm c}$ can be enhanced by one order of magnitude compared to $\overline{T}_{\rm c}$, although the $\gamma$/$\overline{\gamma}$ remains close to one within 30\%.
Instead, this $T_{\rm c}$ enhancement is accompanied by the decrease in $\Theta_{\rm D}$/$\overline{\Theta}_{\rm D}$, which underlines the role of phonon softening in the cocktail effect.
Furthermore, our analysis indicates that the $T_{\rm c}$ mainly controlled by the $\langle\omega_{\rm log}\rangle$ and $\lambda_{\rm ep}$, and the latter increases linearly with the 1/$M_{\rm alloy}$ and is weakened as the increase of $\Delta$$S_{\rm mix}$.
Our results represent a first step toward quantitative understanding the superconducting properties in compositionally complex alloys.

\section*{ACKNOWLEDGEMENT}
We acknowledge financial support by the foundation of Westlake University, the Yunnan Fundamental Research Projects (202201AU070118,202401AT070378), the Analysis and Testing Foundation of Kunming University of Science and Technology (2021T20200152).

\end{document}